\documentclass[amsmath,amssymb,aps,prd,nofootinbib,twocolumn]{revtex4-2}

\usepackage{graphicx}
\usepackage{float}
\usepackage{dcolumn}
\usepackage{bm}
\usepackage{color}
\usepackage[colorlinks=true,linkcolor=blue,citecolor=blue, urlcolor=blue]{hyperref}

\begin{document}

\title{Determination of $|V_{\rm cb}|$ using Bayesian analysis of the $B \rightarrow D^{*}\ell {\bar \nu}_\ell$ semileptonic decay width with four-loop QCD corrections}

\author{Wang Li$^1$}
\email{lwang@stu.cqu.edu.cn}

\author{Xing-Gang Wu$^1$}
\email{wuxg@cqu.edu.cn}

\author{Hua Zhou$^2$}
\email{zhouhua@swust.edu.cn}

\author{Xu-Chang Zheng$^1$}
\email{zhengxc@cqu.edu.cn}

\affiliation{$^1$Department of Physics, Chongqing Key Laboratory for Strongly Coupled Physics,Chongqing University, Chongqing 401331, P.R. China\\
$^2$School of Mathematics and Physics, Southwest University of Science and Technology, Mianyang 621010, P.R. China}


\begin{abstract}

The Cabibbo-Kobayashi-Maskawa matrix element $|V_{\rm cb}|$ is an important Standard Model parameter, whose value can be determined by using the semi-leptonic decay $B\rightarrow D^{*}\ell{\bar \nu}_\ell$. The perturbative QCD (pQCD) corrections to the $B \to D^{*}$ transform form factor ${\cal F}(w)$ has been known up to the N$^3$LO level, whose magnitude remains sensitive to the choice of renormalization scale $\mu_r$. To improve the precision of ${\cal F}(w)$ and hence $|V_{\rm cb}|$, we first apply the single-scale approach of Principle of Maximum Conformality (PMC) to eliminate the conventional (Conv.) renormalization scale dependence of the short-distance parameter $\eta_A$ and then predict the contribution of its unknown N$^4$LO term via Bayesian analysis. In this paper, we adopt two probabilistic models for Bayesian analysis: the Cacciari-Houdeau model (CH model) and the geometric behavior model (GB model). It is shown that by using the PMC series in combination with Bayesian analysis, one can achieve high degree of reliability in estimating unknown higher-order terms. A more convergent behavior is achieved by applying the PMC, confirmed by the predicted N$^4$LO contributions: for the CH model, $\eta_A|_{\rm Conv.}^{\rm N^4LO}=\{-0.0032,+0.0052\}$ and $\eta_A|_{\rm PMC}^{\rm N^4LO}=\{-0.0004,+0.0004\}$; for the GB model, $\eta_A|_{\rm Conv.}^{\rm N^4LO}=\{-0.0049,+0.0075\}$ and $\eta_A|_{\rm PMC}^{\rm N^4LO}=\{-0.0007,+0.0007\}$. Comparing with the latest experimental measurements, we obtain $|V_{\rm cb}||_{\rm PMC}=(40.58^{+0.53}_{-0.57})\times10^{-3}$, which is consistent for both CH and GB models and in good agreement with the PDG world average, $|V_{\rm cb}|_{\rm PDG}=(41.1\pm1.2)\times10^{-3}$ within errors.

\end{abstract}

\maketitle

\section{Introduction}

The Cabibbo-Kobayashi-Maskawa (CKM) matrix elements are essential components of the Standard Model (SM), describing the mixing of quark flavors~\cite{Matsuoka:1998xt}. Their precise determination is crucial for testing the SM's validity and deepening our understanding of flavor structure. In particular, a precise determination of the CKM matrix element $|V_{\rm cb}|$ is vital for reducing the uncertainties of the unitary triangle and improving predictive accuracy in flavor physics -- especially for indirect CP violation, which is highly sensitive to new physics signals. Among these, the semileptonic decay process $B{\rightarrow}D^{*}\ell{\bar \nu}_\ell$ ($\ell=e,\mu$) is pivotal for precise extraction of $|V_{\rm cb}|$. In recent years, both the LHC and B factories have provided increasingly precise measurements of the differential decay width and related observables for the $B{\rightarrow}D^{*}\ell{\bar \nu}_\ell$ process~\cite{ALEPH:1995qbg, Belle:2001gkd, ALEPH:1996dlo, Belle:2001nft, Jaiswal:2017rve, OPAL:1996vml}. Theoretically, the differential decay width of the exclusive process $B \rightarrow D^{*}\ell{\bar \nu}_\ell$ can be expressed as~\cite{Ricciardi:2019zph}
\begin{eqnarray}
\frac{d\Gamma}{dw}(B\to D^*\ell{\bar \nu}_\ell)&=&\frac{G_F^2}{48\pi^3}(m_B-m_{D^*})^2m_{D^*}^3\nonumber
\\&\times& \chi(w)(w^2-1)^{\frac12}|V_{\rm cb}|^2\nonumber
\\&\times& |\eta_{\rm EW}|^2|{\cal F}(w)|^2,
\label{eq1}
\end{eqnarray}
where $G_F$ is the Fermi constant, and $m_{B,D^{\ast}}$ denotes the mass of the $B$ and $D^{\ast}$ meson, respectively. The recoil parameter $w$ is defined as $w = (m_{B}^{2}+m_{D^{*}}^{2} - q^{2})/(2m_{B}m_{D^{*}})$. The term $|\eta_{\rm EW}|$ refers to the electroweak correction factor, and ${\cal F}(w)$ represents the transition form factor associated with this process. While $\chi(w)$ denotes an additional phase-space factor, whose expression is~\cite{Ricciardi:2019zph}:
\begin{eqnarray}
\chi(w)=&& (w+1)^2\nonumber \\
 && \times \left(1+\frac{4w}{w+1}\frac{m_{B}^{2}-2w m_{B}m_{D^{*}}+m_{D^{*}}^2 }{(m_{B}-m_{D^{*}})^2} \right).
\end{eqnarray}
The parameter $|{V}_{\rm cb}|$, an important input to Eq.(\ref{eq1}), can be extracted by comparing the experimentally measured differential decay width with the theoretically calculated transition form factor ${\cal F}(w)$. Therefore, a precise determination of $|{\cal F}(w)|$ is crucial for the accurate extraction of $|{V}_{\rm cb}|$. Using the QCD factorization method, ${\cal F}(w)$ can be expressed as ${\cal F}(w)=\eta_A \hat{\xi}(w)$, where $\eta_A$ and $\hat{\xi}(w)$ represent the perturbative short-distance coefficient and long-distance hadronic dynamics~\cite{Neubert:1994vy}, respectively. In heavy quark effective theory~\cite{Manohar:2000dt, Luke:1990eg}, the long-distance hadronic dynamics $\hat{\xi}(w)$ corresponds to the Isgur-Wise function, normalized such that $\hat{\xi}(1)=1+\delta_{1/m^2_Q}$, where $Q$ denotes a heavy quark ($b$ or $c$). 

Currently, the short-distance coefficient $\eta_A$ has been calculated up to next-to-next-to-next-to-leading order (N$^3$LO) QCD corrections~\cite{Archambault:2004zs}. However, the N$^3$LO series still exhibits significant renormalization scale ($\mu_r$) dependence. Conventionally, $\mu_r$ is set to a typical value (e.g., $Q$) to suppress large logarithmic terms in the expansion coefficients and then varied within a range, e.g., $[Q/n, nQ]$ (where $n = 2, 3, 4, \cdots$), to estimate theoretical uncertainties. This scale-setting procedure is inherently arbitrary, leading to a mismatch between the perturbative expansion and the strong coupling constant $\alpha_s(\mu_r)$ and causing the pQCD series to depend on the chosen $\mu_r$. Moreover, this renormalization scale uncertainty violates the requirements of renormalization group invariance (RGI) and degrades the accuracy of theoretical predictions.

The Principle of Maximum Conformality (PMC) offers a process-independent method to eliminate conventional renormalization-scale ambiguities and reveal the perturbative nature of the QCD series~\cite{Brodsky:2011ig, Brodsky:2011ta, Mojaza:2012mf, Brodsky:2012rj, Brodsky:2013vpa}. The scale-running behavior of $\alpha_s$ is governed by the renormalization group equation (RGE). Leveraging this, the PMC approach systematically absorbs all RGE-involved {$\{\beta_i\}$}-terms of the series to determine the accurate magnitude of the strong coupling, ensuring a consistent fixed-order pQCD expansion~\footnote{If there exist additional $\{\beta_i\}$-terms associated with other scale-dependent parameters (e.g., parton distribution functions and running quark masses), these terms should be retained as ``conformal" ones when determining the magnitude of $\alpha_s$ but will be utilized to fix the appropriate magnitudes of these parameters via their respective RGEs~\cite{Yan:2024oyb}.}. The resulting perturbative coefficients become $\mu_r$-independent and exhibit conformal properties. Notably, the PMC-improved series, with accurate $\alpha_s$ value, is also renormalization-scheme invariant~\cite{Wu:2014iba, Wu:2015rga, Wu:2019mky, Yan:2023hra}, which is also ensured by the commensurate scale relations among the pQCD approximants under different schemes~\cite{Brodsky:1994eh}.

The $\mu_r$-independent PMC series also facilitates the estimation of unknown higher-order (UHO) terms~\cite{Du:2018dma, Shen:2022nyr, Wu:2025xdv}. In the 1990s, the Pad$\acute{e}$ approximation approach (PAA) was used to calculate UHO terms~\cite{Samuel:1995jc, Samuel:1992qg}, where PAA employs the rational generating function to approximate the known finite-term power series and infers UHO terms via series expansion. Ref.~\cite{Zhou:2022fyb} utilized a [0,2]-type generating function to estimate the N$^4$LO term for $\eta_A$, yielding a highly accurate prediction of $|V_{\rm cb}|$. As another innovative attempt, Bayesian analysis (BA) has emerged as an efficient method for estimating UHO contributions~\cite{Cacciari:2011ze, Bonvini:2020xeo, Duhr:2021mfd}. BA calculates UHO terms using probability density distributions, specifically by constructing probability distributions and iteratively updating probabilities as new information becomes available. The BA approach has been successfully applied in various high-energy processes~\cite{Luo:2023cpa, Shen:2023qgz, Yan:2024hbz}. In this paper, we use conformal coefficients obtained via the PMC single-scale approach (PMCs)~\cite{Shen:2017pdu, Yan:2022foz} and apply BA to estimate the N$^4$LO term of $\eta_A$, then incorporate the improved high-precision $\eta_A$ into the extraction of $|V_{\rm cb}|$.

The rest of the paper is organized as follows. In Sec.\ref{sec2}, we sketch the methods used in the calculation, e.g., the PMCs approach and the Bayesian approach. In Sec.\ref{sec3}, we present the numerical results and discussions. Sec.\ref{sec4} is reserved as a summary.

\section{Calculation method}
\label{sec2}

\subsection{The principle of maximum conformality}

The perturbative expression of $\eta_A$ up to $\rm N^3LO$-level QCD corrections can be written as:
\begin{eqnarray}
\eta_A &=&1+\sum^{3}_{i=1}r_{i}a^i_s(\mu_r),
\end{eqnarray}
where $a_s=\alpha_s/\pi$. This perturbative series is commonly calculated under the $\overline{\rm MS}$ scheme (the scheme used to define $\alpha_s$), e.g., Ref.~\cite{Archambault:2004zs}. However, as observed in Ref.~\cite{Zhou:2022fyb}, the physical V scheme~\cite{Appelquist:1977tw,Fischler:1977yf,Peter:1996ig,Schroder:1998vy}, which is defined in the static potential between a heavy quark and a heavy antiquark, is more effective than the $\overline{\rm MS}$ scheme for $\eta_A$. In the V scheme, the small scale problem can be avoided and the convergence of this perturbative series can be improved. Thus, in the following calculations, we work in the V scheme. The expansion coefficients $r_{i}$ under the V scheme can be found in Ref.~\cite{Zhou:2022fyb}.

By using the QCD degeneracy relations~\cite{Bi:2015wea}, the expansion coefficients $r_i$ can be decomposed into conformal and non-conformal parts:
\begin{eqnarray}
	r_1&=&r_{1,0},\\
	r_2&=&r_{2,0}+\beta_0 r_{2,1},\\
	r_3&=&r_{3,0}+\beta_{1}r_{2,1}+2\beta_{0}r_{3,1}+\beta^{2}_{0}r_{3,2}.
\end{eqnarray}
Based on RGE, the PMCs resums all known types of $\{\beta_i\}$-terms and determines an overall effective value of $\alpha_s$ (and hence its PMC scale). According to the standard PMCs procedure~\cite{Shen:2017pdu, Yan:2022foz} the pQCD approximant of $\eta_A$ becomes free of RGE-involved $\{\beta_i\}$-terms:
\begin{eqnarray}
\eta_{A|\rm PMCs}&=&1+\sum^{3}_{i=1}{r}_{i,0} a^{i}_{s}(Q_\ast),
\end{eqnarray}
where $r_{i,0}$ are conformal coefficients and the PMC scale $Q_\ast$ of the $B \rightarrow D^{*}\ell{\bar \nu}_\ell$ decay can be determined up to next-to-leading logarithm (NLL) accuracy,
\begin{eqnarray}
\ln\left(\frac{Q^{2}_{\ast}}{Q^2}\right) &=& T_{0} + T_{1} a_s(Q),
\label{equ1}
\end{eqnarray}
where $Q$ represents the typical momentum flow of the process, which is usually set as $\sqrt{m_b m_c}$ for the present process. The explicit expressions for the expansion coefficients $r_{i}$, $r_{i,0}$ and $T_{i}$ can be found in the Ref.~\cite{Zhou:2022fyb}.

\subsection{The Bayesian analysis approach}

We will estimate the contribution of the unknown $\rm N^{4}LO$ term by using the probability distribution, i.e., the BA approach, to derive a more accurate pQCD result. The BA approach serves as an effective approach for constructing probability distributions, founded on the Bayesian theorem and using known information about the perturbative series to create a probability model. When the convergence of the perturbation series is good, the BA approach has high efficiency and predicts reasonable UHO contribution even by using lower fixed-order series. In the paper, we will adopt two Bayesian models, namely the Cacciari-Houdeau (CH) model~\cite{Cacciari:2011ze} and the geometric behavior (GB) model~\cite{Bonvini:2020xeo}, to analyze the contribution of the unknown $\rm N^{4}LO$ term to $\eta_A$ and discuss their uncertainties in detail.

\subsubsection{CH model}

The CH model is one of the earliest probabilistic models developed in 2010 for predicting UHO terms, it stands out as one of the most understandable and widely adopted models. A physical quantity that has been calculated up to order $a_s^n$ can be written as:
 \begin{eqnarray}
 \rho_n=\sum_{i=0}^n c_i a_s^i.
 \label{eq.rho-n}
 \end{eqnarray}
 
 There are three hypotheses in the CH model:
 \begin{itemize}
 \item The model assumes that all the coefficients $c_i$ of the perturbative series are bounded in absolute value by a common parameter $\bar{c}$. The logarithm of $\bar{c}$ is described by a flat probability distribution over the entire parameter space, i.e.,
\begin{eqnarray}
g({\rm ln}\, \bar{c})=\frac{1}{2|\ln\epsilon|}\theta\left(\vert {\rm ln} \epsilon \vert  -\vert {\rm ln} \bar{c} \vert\right)\label{e1},
\end{eqnarray}
where $\theta(x)$ is the Heaviside step function, and the parameter $\epsilon$ is taken to the limit $\epsilon\rightarrow0$ at the end. Here, assigning a flat distribution to ${\rm ln}\, \bar{c}$ reflects the absence of prior knowledge about the order of magnitude of $\bar{c}$. Then such distribution can be expressed as the probability distribution of $\bar{c}$:
\begin{eqnarray}
g_0(\bar{c})=\frac{1}{2|\ln\epsilon|}\frac{1}{\bar{c}} \theta\left(\frac{1}{\epsilon}-\bar{c}\right)\theta(\bar{c}-\epsilon)\label{e2}.
\end{eqnarray}

\item With the parameter $\bar{c}$ known, the conditional probability for an unknown coefficient $c_k$ -- referred to as the likelihood probability -- is given as a uniform distribution in the CH model, i.e.,
\begin{eqnarray}
h_0(c_k|\bar{c})=\frac{1}{2\bar{c}}\theta(\bar{c}-|c_k|).
\label{e3}
\end{eqnarray}

\item The model also assumes that the coefficients are independent from each other, which implies that
\begin{eqnarray}
h(c_i, c_j|\bar{c})=h_0(c_i|\bar{c})h_0(c_j|\bar{c}).
\label{e4}
\end{eqnarray}
 \end{itemize}

Using the Bayesian conditional probability formula, the probability distribution for an uncalculated coefficient $c_k$ is
\begin{eqnarray}
f_c(c_{k}|c_{0},\ldots,c_{n})&=&\frac{f(c_{k},c_{0},\ldots,c_{n})}{f(c_{0},\ldots,c_{n})}(k>n)\nonumber \\
 &=&\frac{\int d\bar{c}\, h_0(c_{k}|\bar{c})h_0(c_{0}|\bar{c})\cdots h_0(c_{n}|\bar{c})g_{0}(\bar{c})}{\int d\bar{c}\, h_0(c_{0}|\bar{c})\cdots h_0(c_{n}|\bar{c})g_{0}(\bar{c})}.\nonumber
 \\
\label{e5}
\end{eqnarray}
Inserting Eq.(\ref{e2}) and Eq.(\ref{e3}) into Eq.(\ref{e5}), one obtains
\begin{eqnarray}
f_c(c_{k}|c_{0},\ldots,c_{n})=\left\{\begin{array}{cc}\frac{(n+1)}{2(n+2)\bar{c}_{(n)}}, &|c_k|\leq\bar{c}_{(n)}\\\\\frac{(n+1)\bar{c}_{(n)}^{(n+1)}}{2(n+2)|c_k|^{n+2}}, &|c_k|>\bar{c}_{(n)}\end{array}\right.
\label{e6}
\end{eqnarray}
where $c_0,\cdots,c_n$ are the known coefficients of the pertubative series, and $\bar{c}_{(n)}=max(|c_0|,\cdots,|c_n|)$. It is observed that Eq.(\ref{e4}) is symmetric about the origin, this indicates that we are generally unable to know the positive or negative nature of the UHO terms. 

In Bayesian analysis, one can use the probability distribution equation described above to calculate $c_k$ at a specific degree of confidence, referred to as degree-of-belief (DoB). This interval is known as the credible interval (CI), and its corresponding definition is given by~\cite{Shen:2022nyr}:
\begin{eqnarray}
p\%&=&\int_{-c_k^{(p)}}^{c_k^{(p)}}f_c(c_k|c_0,\ldots,c_n) \mathrm{d}c_k,
\end{eqnarray}
and the distribution interval for the unknown coefficients $c_k$ with a specific DoB can be determined:
\begin{eqnarray}
c_k^{(p)}&=&\begin{cases}\bar{c}_{(n)}\frac{n+2}{n+1} p\%,&p\%\leq\frac{n+1}{n+2}\\ \bar{c}_{(n)}\left[(n+2)(1-p\%)\right]^{-\frac{1}{n+1}} ,&p\%>\frac{n+1}{n+2}\end{cases}.
\end{eqnarray} 

\subsubsection{GB model}

Another Bayesian model adopted in this paper is the GB model~\cite{Bonvini:2020xeo}. In the GB model, a standard perturbative series up to order $a_s^n$ is expressed as follows:
\begin{eqnarray}
\rho_n =\rho_0\sum_{k=0}^n\delta_k,
\end{eqnarray}
where
\begin{eqnarray}
\delta_{k}\equiv\frac{\rho_{k}-\rho_{k-1}}{\rho_{0}},
\end{eqnarray} 	
denotes the ratio of the difference between the $(k+1)$-term series and the $k$-term seires to the first term of the series. 

The fundamental assumptions of this model are:
\begin{itemize}
\item The terms $\delta_k$ are bounded by
\begin{eqnarray}
|\delta_k|\leq c\,a^k,
\end{eqnarray}
where symbols $c$ and $a$ are hidden parameters of the model that do not appear in the final calculations. Unlike the CH model, the assumption of the GB model does not incorporate the information of the coupling constant $\alpha_s$, instead, it is included in the $\delta_k$. In the GB model, the corresponding prior probability distribution for $c$ is given by:
\begin{eqnarray}
g_{0}(c)=\frac{\epsilon}{c^{1+\epsilon}}\theta(c-1),\label{e7}
\end{eqnarray}
where $\epsilon$ is typically a very small positive value. The prior probability distribution of $a$ is given by:
\begin{eqnarray}
g_0(a)=(1+\omega)(1-a)^\omega\theta(a)\theta(1-a),\label{e8}
\end{eqnarray}
where $\omega$ is an integer greater than 1. 

\item If the probability distribution of $c$ and $a$ are known, the probability distribution for a term $\delta_k$ is assumed as:
\begin{eqnarray}
h_0(\delta_{k}|c,a)=\frac{1}{2c\,a^{k}}\theta\left(c\,a^{k}-|\delta_{k}|\right).\label{e9}
\end{eqnarray}

\item The terms $\delta_k$ are independent from each other, i.e.,
\begin{eqnarray}
h(\delta_i, \delta_j|c,a)=h_0(\delta_i|c,a)h_0(\delta_j|c,a).
\label{e10}
\end{eqnarray}

\end{itemize}

Using Bayesian theorem, one can obtain the distribution of the first UHO term:
\begin{eqnarray}
f_{\delta}(\delta_{n+1}|\delta_{n},\ldots,\delta_{1})=\frac{f(\delta_{n+1},\delta_{n},\ldots,\delta_{1})}{f(\delta_{n},\ldots,\delta_{1})},
\label{e11}
\end{eqnarray}
where
\begin{eqnarray}
&&f(\delta_{m},\ldots,\delta_{1})\nonumber\\=&&\int dc\int da\, h(\delta_{m},\ldots,\delta_{1}|c,a)g_{0}(c)g_{0}(a).\label{e12}
\end{eqnarray}
Inserting Eqs.(\ref{e7}-\ref{e10}) into Eq.(\ref{e11}) yields its probability distribution. The calculation based on the GB model can be carried out using the program THunc~\cite{Bonvini:2020xeo}.

Having the probability distribution for UHO terms, the DoB is
\begin{eqnarray}
p\%&=&\int_{-\delta_{n+1}^{(p)}}^{\delta_{n+1}^{(p)}}f_{\delta}(\delta_{n+1}|\delta_1,\ldots,\delta_n) \mathrm{d}\delta_{n+1}.
\end{eqnarray}
Then, the CI $\delta_{n+1}^{(p)}$ under a specific DoB can be determined using this formula.

\section{Numerical results and discussions}
\label{sec3}

In this section, we present the numerical results for $\eta_A$ and $|V_{\rm cb}|$. In the numerical calculation, the pole masses are taken as $m_c=1.68$ GeV and $m_b=4.78$ GeV, and the QCD asymptotic scale parameter is taken as $\Lambda^V_{{\rm QCD}{| n_f=4}}=283.0^{+16.1}_{-15.6}{\rm MeV}$~\footnote{The error in $\Lambda^V_{{\rm QCD}| n_f=4}$ originates from the uncertainty in the strong coupling constant under the $\overline{\rm MS}$ scheme, $\alpha_s(M_{_Z})=0.1179\pm0.0009$~\cite{ParticleDataGroup:2022pth}. Specifically, the $\overline{\rm MS}$-scheme uncertainty in $\alpha_s(M_{_Z})$ yields $\Lambda^{\overline{\rm MS}}_{{\rm QCD}| n_f=4}=207.2^{+11.8}_{-11.4}{\rm MeV}$, which in turn propagates to $\Lambda^V_{{\rm QCD}| n_f=4}=283.0^{+16.1}_{-15.6}{\rm MeV}$.} for the V-scheme strong coupling constant~\cite{Zhou:2022fyb}. 

The numerical results for $\eta_A$ up to $\rm N^{3}LO$ QCD corrections under the conventional (Conv.) and PMCs approaches are
\begin{eqnarray}
 \eta_{A}|_{\rm{Conv.}}^{\mu_{r}=Q/2}&=&\{1, -0.0953,\ \ \,0.0921, -0.0259\}, \\
 \eta_{A}|_{\rm{Conv.}}^{\mu_{r}=Q}&=&\{1, -0.0625, -0.0093, -0.0017\},  \\
 \eta_{A}|_{\rm{Conv.}}^{\mu_{r}=2Q}&=&\{1, -0.0475, -0.0336, -0.0319\},  \\
 \eta_{A}|_{\rm{PMCs.}} &=&\{1, -0.0644, -0.0158, \ \ \,0.0027\}.
\end{eqnarray}
In conventional scale-setting approach, the central value of $\mu_r$ is set as $Q=\sqrt{m_{b}m_{c}}$. Then, the variation of $\mu_r$ within the range $[Q/2, 2Q]$ sets the upper and lower bounds of the theoretical error, leading to significant $\mu_r$ uncertainty. It is found that the conventional series does show strong scale dependence, leading to quite different perturbative behaviors. In contrast, the PMC approach systematically removes the conventional renormalization scale uncertainty, and the value of $\eta_A$ remains invariant under renormalization scale variations. This enables the scale-invariant PMC conformal series to more effectively estimate the contribution of UHO terms.

\subsection{The estimated magnitude of unknown N$^4$LO-terms using PAA}

In the literature, the PAA has been employed to estimate the contributions of UHO terms. For a comparison, we first present the estimate of unknown N$^4$LO-terms based on the PAA. For a perturbative series $\rho_n$, e.g., given in Eq.(\ref{eq.rho-n}), the approximation function of PAA can be expressed as follows~\cite{Samuel:1995jc}:
 \begin{eqnarray}
 \rho_n^{[N/M]}&=&\frac{p_0+p_1\,a_s+\cdots+p_N\,a_s^N}{1+q_1\, a_s +\cdots q_M\,a_s^M},
 \label{eq.PAA}
 \end{eqnarray}
where $N$ and $M$ (with $N + M = n$) denote the highest powers of the numerator and denominator polynomials, respectively, while $p_i$ and $q_i$ are input parameters. By comparing the Taylor expansion series of fractional generating function (\ref{eq.PAA}) with the original series in Eq.(\ref{eq.rho-n}) up to order $a_s^n$, the coefficients $p_i$ and $q_i$ can be uniquely determined via the given series. Subsequently, one can expand Eq.(\ref{eq.PAA}) to a higher order, e.g., order $a_s^{n+1}$, to have an estimation of the uncalculated $a_s^{n+1}$ term.

\begin{table}[htb]
\renewcommand{\arraystretch}{1.5}
\setlength{\tabcolsep}{12pt}
\begin{tabular}{c c c  c}
\hline
~ &$\eta_{A}|_{\rm{Conv.}}^{\rm{N^{4}LO}}$& $\eta_{A}|_{\rm {PMCs}}^{\rm{N^{4}LO}}$\\
\hline
$[0/3]$-type & $0.0044_{+0.0040}^{+0.0106}$& 0.0001\\
 $[1/2]$-type & $-0.0301_{-0.0290}^{+0.0123}$& 0.0001\\
 $[2/1]$-type & $-0.0003_{-0.0299}^{+0.0075}$& -0.0005\\
\hline
\end{tabular}
\caption{The predictions for the $\rm N^4LO$ term of $\eta_A$, estimated using the three types of PAA under both the conventional and PMCs approaches, show that the error in the conventional series arises from the renormalization scale with $\mu_{r}\in[Q/2,2Q]$. }
\label{t2}
\end{table}

Up to now, the perturbative series of $\eta_A$ is known up to $\rm N^{3}LO$-level; there are three types of PAA, namely [0/3], [1/2], and [2/1]. Uncertainties of different $[N/M]$-types can be treated as the systematic errors of the PAA method itself. The predicted $\rm N^{4}LO$ coefficient $c_4$ by the three types are as follows:
\begin{eqnarray}
 &&c_4^{[0/3]}=\frac{c_1^4-3 c_0\, c_1^2\, c_2+c_0^2\, c_2^2+2c_0^2\, c_1\, c_3}{c_0^3},
 \\
 &&c_4^{[1/2]}=\frac{-c_{2}^{3}+2c_{1}\,c_{2}\,c_{3}-c_{0}\,c_{3}^{2}}{c_{1}^{2}-c_{0}\,c_{2}},
 \\
 &&c_4^{[2/1]}=\frac{c_3^2}{c_2}.
\end{eqnarray}
The estimated values for the $\rm N^4LO$ term of $\eta_A$ are presented in Table \ref{t2}. The conventional results indicate that different types of PAA result in significant variations of $\eta_A|^{\rm N^{4}LO}_{\rm Conv.}$. In Table \ref{t2}, the error of the conventional results is caused by taking $\mu_{r}\in [Q/2,2Q]$, whose magnitude is $(^{+241\%}_{+91\%})$ for [0/3]-type, $(^{-41\%}_{+96\%})$ for [1/2]-type, and $(^{+250\%}_{-9967\%})$ for [2/1]-type, respectively. For the subsequent discussion, we will use the squared average of the results of the three types as the final prediction. Compared to the conventional scale-setting approach, the PMCs not only substantially reduces the $\mu_r$ uncertainty but also achieves consistent predictions across three distinct types of PAA. This significantly enhances the predictive accuracy of the $\rm N^{4}LO$ term. 

\subsection{The estimated magnitude of unknown N$^4$LO-terms using CH model of BA}

\begin{table*}[htb]
\renewcommand\arraystretch{1.5}
\setlength{\tabcolsep}{12pt}
\begin{tabular}{c c c c c}
\hline
&$c_1$&$c_2$&$c_3$&$c_4$\\ 
\hline
Conv.	&$[-10.0000,+10.0000]$&$[-2.5820,+2.5820]$&$[-1.8064,+1.8064]$&$[-2.9504,+2.9504]$\\
PMCs	&$[-10.0000,+10.0000]$&$[-2.5820,+2.5820]$&$[-2.8989,+2.8989]$&$[-4.2332,+4.2332]$\\
\hline
\end{tabular}
\caption{The CI (${\rm DoB}=95\%$) for the coefficients $c_i$ ($\mu_r$) ($i = 1,2,3,4$) at the scale $\mu_r = Q$ under the conventional approach and the scale-invariant coefficients $c_i$ under PMCs approach.}
\label{t3}
\end{table*}

We now present the numerical results based on the two BA models in terms of a probability distribution. Applying the CH model to $\eta_A$ with the $95\%$ DoB, we obtain the estimate for the CI of higher order coefficients. The estimated cofficents are presented in Table \ref{t3}. The results for $c_i$ are estimated based on the known coefficients ($c_0, c_1, \cdots, c_{i-1}$). These intervals for coefficients are symmetric about the zero. 

\begin{figure}[h]
\includegraphics[width=0.45\textwidth]{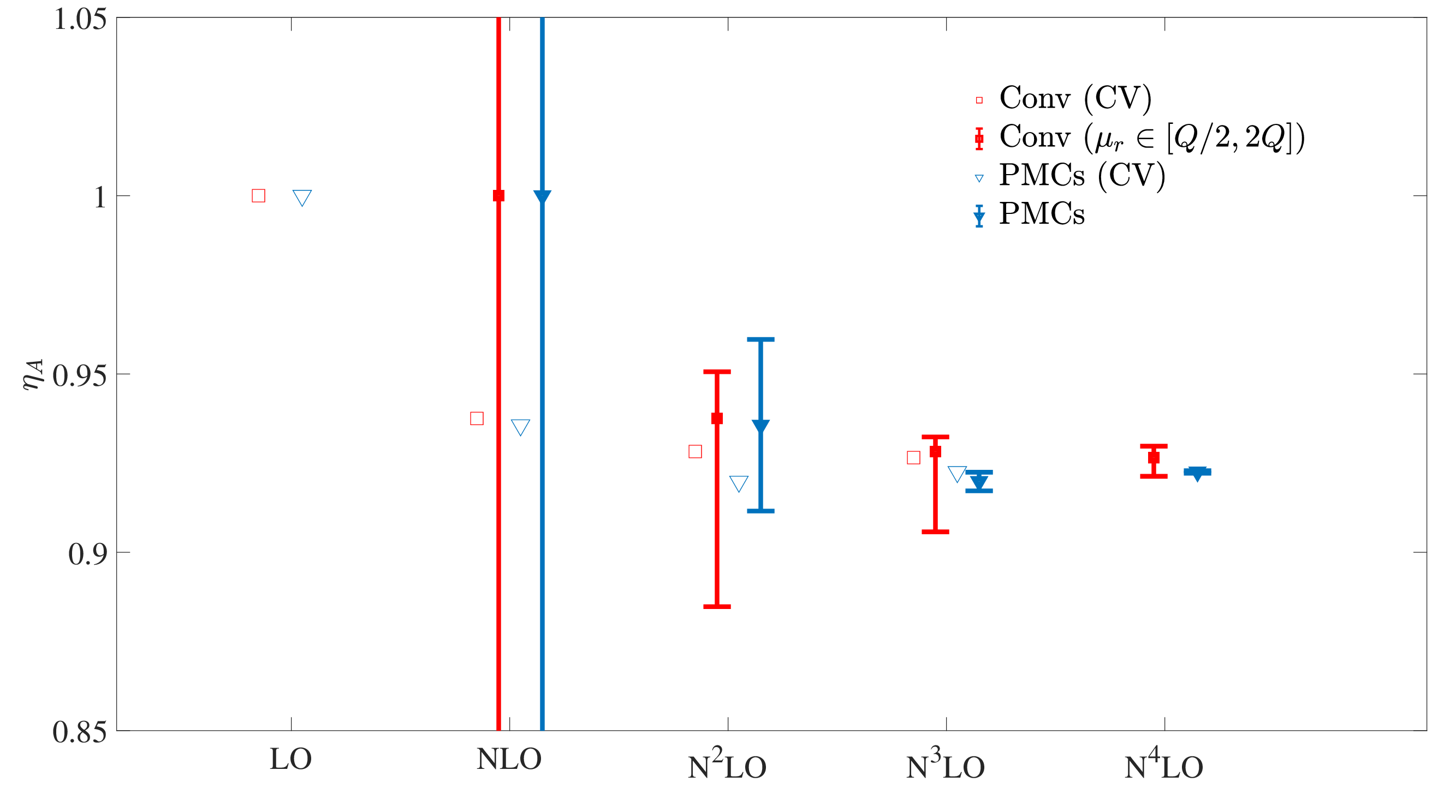}
\caption{Comparison of the calculated central values of $\eta_A$ for the known series (labeled as “CV”) with the predicted CIs (with DoB=95\%) of $\eta_A$ up to $\rm N^{4}LO$ level. The red hollow squares and blue hollow triangles represent the calculated central values of the known fixed-order pQCD predictions under conventional and PMCs approaches, respectively. The red square, blue triangle, and their corresponding error bars represent the BA (CH model) predictions for $\eta_A$ under conventional and PMC approaches, respectively.}
 \label{fig:epsart1}
\end{figure}

In Fig.\ref{fig:epsart1}, we present a comparison between the central values of $\eta_A$ calculated using the known series and the CIs (with DoB=95\%) of $\eta_A$ predicted based on the CH model. The predicted $\eta_A$ at the NLO level falls within the range of an interval but carries significant uncertainty, this is due to the fact that we have less known information about the coefficients when $n=0$. With the increase of order, the CI of $\eta_A$ is greatly reduced. For example, compared to the estimated CI at the $\rm N^{3}LO$ level, the estimated CI of $\eta_A$ at the $\rm N^{4}LO$ level is reduced by a factor of four. Furthermore, the scale dependence ($\mu_{r}\in[Q/2,2Q]$) of the conventional approach leads to a widening of the coefficient prediction interval, resulting in a large uncertainty for $\eta_A$. On the contrary, the uncertainties of the PMCs results are much smaller, e.g., the CI of $\eta_A$ at the $\rm N^{4}LO$ level almost converges to one point. Therefore, the PMCs approach greatly improves the accuracy of the prediction to the $\rm N^{4}LO$ term. In Table \ref {t4}, we also provide detailed numerical results for the $\rm N^{4}LO$ term of $\eta_A$ for further illustration.
\begin{table}
\renewcommand\arraystretch{1.5}
\begin{tabular}{c c}
\hline
 		&	The predicted $\rm N^4LO$ term of $\eta_A$ under the CH model\\ 
\hline
 		Conv.	&$[-0.0032,+0.0052]$\\
 		PMCs	&$[-0.0004,+0.0004]$\\
\hline
\end{tabular}
\caption{The predicted $\rm N^4LO$ term of $\eta_A $ using the CH model under the PMCs approach and conventional approach, where the initial renormalization scale is taken as the range $\mu_{r}\in[Q/2,2Q]$.}
\label{t4}
\end{table}
 
\begin{figure}[h]
\includegraphics[width=0.45\textwidth]{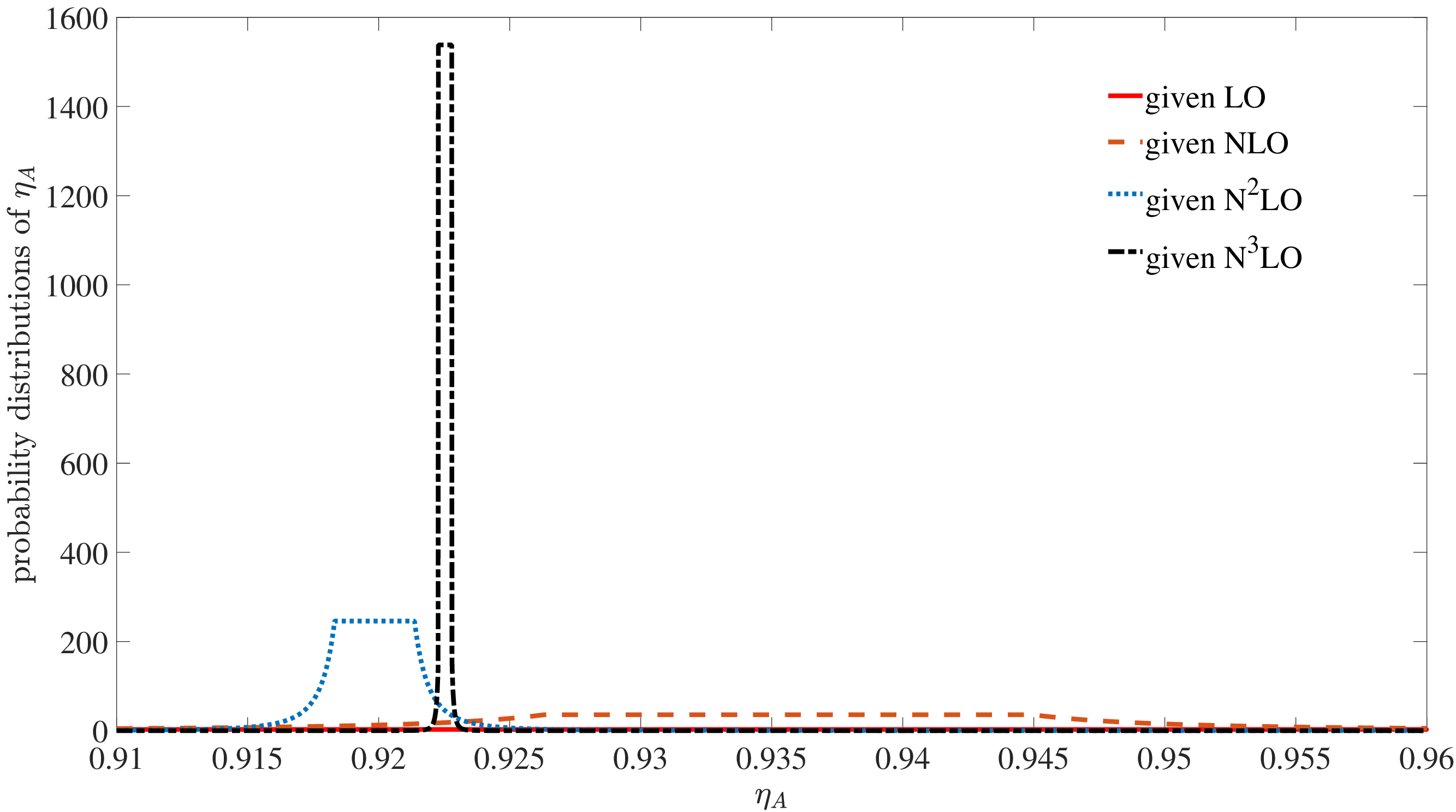}
\caption{The probability density distributions of $\eta_A$ with the PMCs approach under the CH model. The solid line, dashed line, dotted line, dotted-dashed line are results for the given LO, $\rm NLO$, $\rm N^2LO$, and $\rm N^3LO$ series, respectively.}
\label{fig:epsart2}
\end{figure}

We also present the probability density distribution of $\eta_A$ after using the PMCs approach and the CH model in Fig.\ref{fig:epsart2}. The four lines in Fig.\ref{fig:epsart2} correspond to different levels: given LO (solid line), given NLO (dashed line), given $\rm N^2LO$ (dotted line), and given $\rm N^3LO$ (dotted-dashed line). The prominent shapes of those curves indicate the highest possibility of the $\eta_A$ values. Their shapes resemble a towering flat straight line at the center, with suppressed tail curves trailing on each side, resulting in an overall symmetric graph. Fig.\ref{fig:epsart2} shows that as the known information is updated, the probability of the straight-line portion becomes larger while its interval width decreases, tending to converge to one point. Simultaneously, the probability of the curved portions on each side gradually converges to zero. 

\subsection{The estimated magnitude of unknown N$^4$LO-terms using GB model of BA}

\begin{figure}[ht]
	\includegraphics[width=0.45\textwidth]{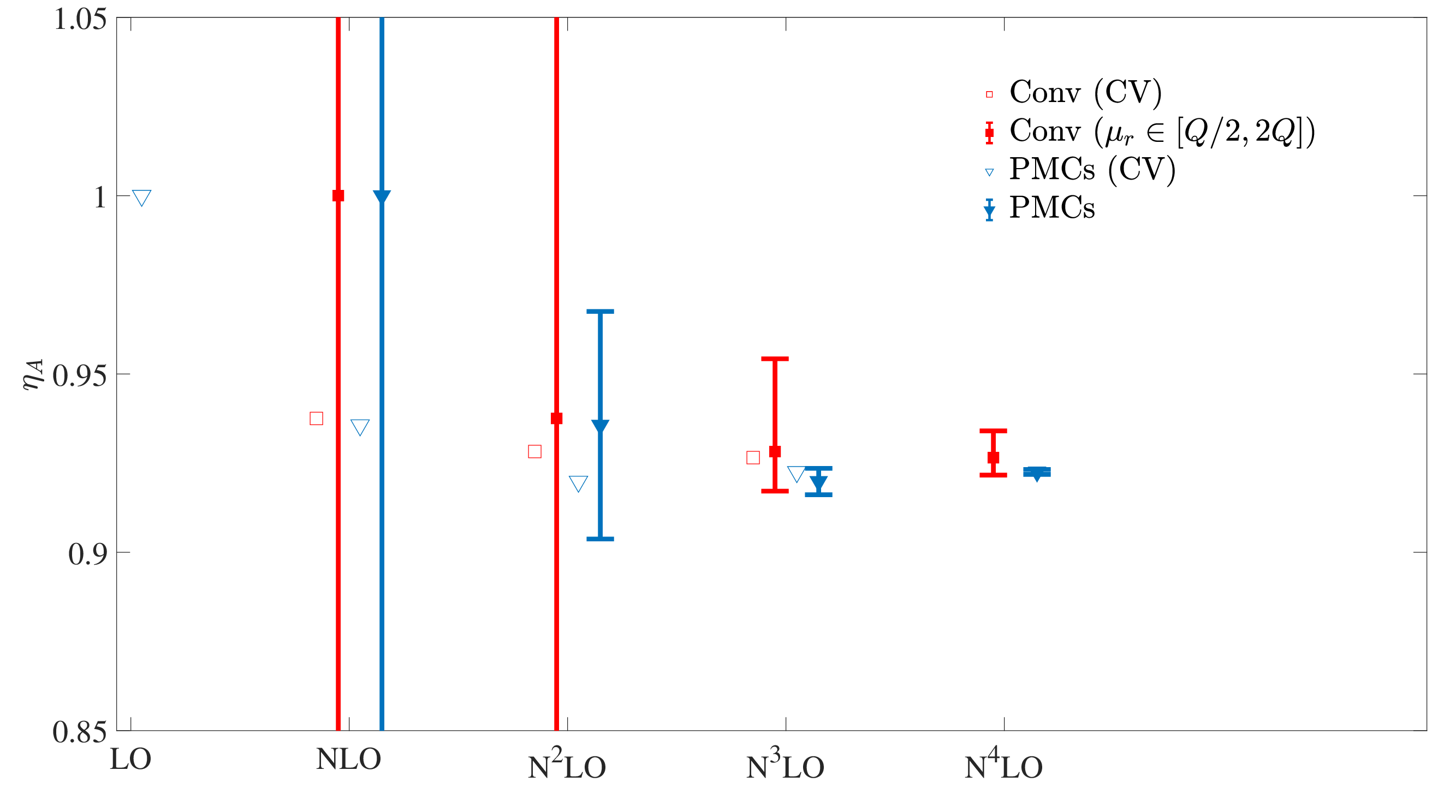}
	\caption{Under the GB model, the predicted CI with the typical $95\%$ DoB for the $\eta_A$ by the conventional scale-setting approach (where the renormalization scale $\mu_{r}\in[Q/2,2Q]$) and PMCs approach, respectively.}
	\label{fig:epsart3}
\end{figure}
 	
By using the GB model with the parameters $\epsilon = 0.01$ and $ \omega = 2$ to the perturbative series of $\eta_A$, we obtain its UHO contributions, which are depicted in Fig.\ref{fig:epsart3}. The figure shows that the results from the conventional scale-setting approach bring large renormalization scale uncertainty. In contrast, the results from the PMCs approach do not have this uncertainty. It can be observed that, as the amount of known perturbative information increases, the CIs (${\rm DoB}=95\%$) significantly decrease. Additionally, the results from the $\rm N^4LO$ term indicate that the prediction intervals nearly overlap at certain point, aligning well with the characteristics of the GB model. It is notable that the results from each order are encapsulated within the results from the preceding order, demonstrating the strong convergence of the GB model. 

\begin{figure}[htb]
\includegraphics[width=0.45\textwidth]{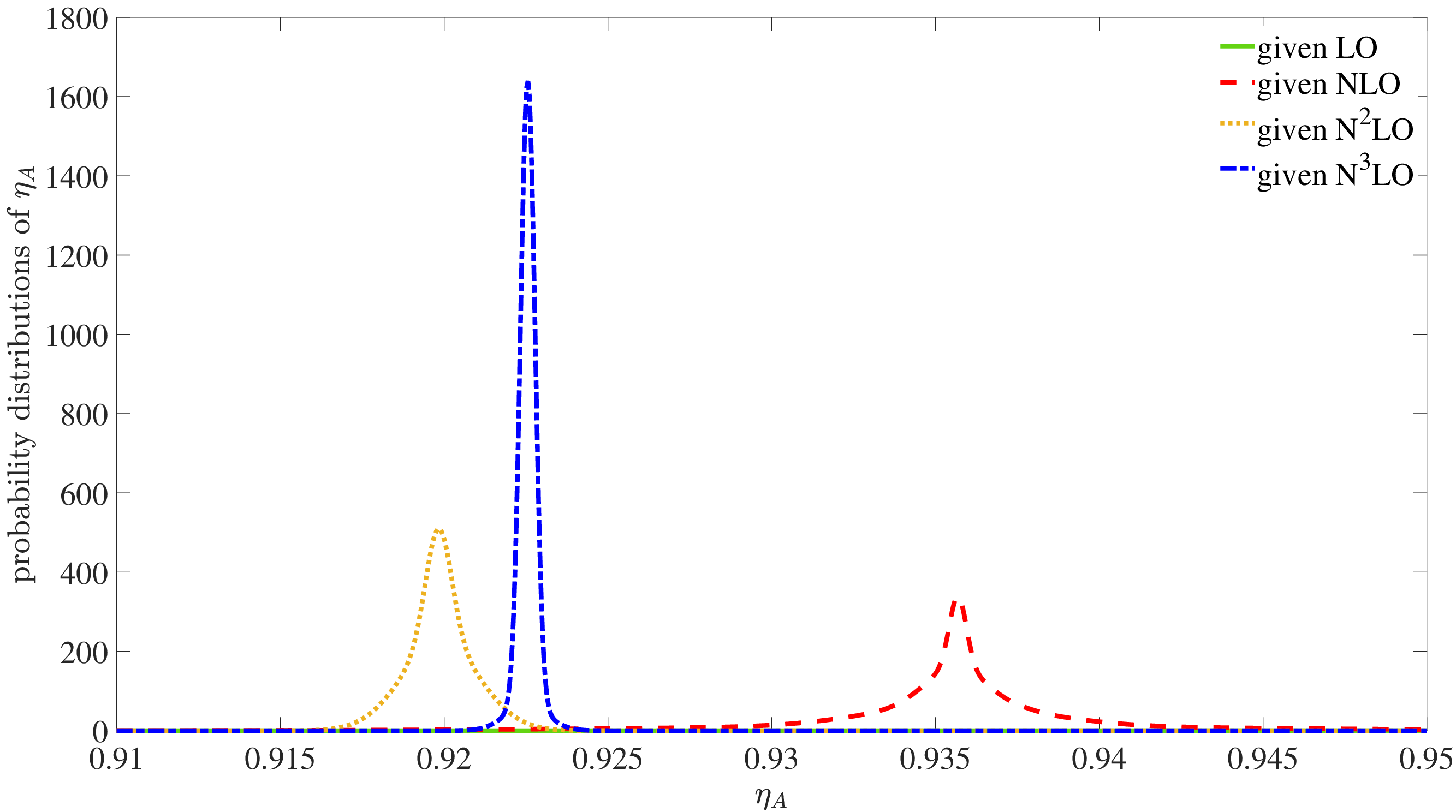}
\caption{The probability density distributions of $\eta_A$ with different orders of knowledge predicted by PMCs under the GB model. The solid line, long dashed line, dotted line and dashed-dotted line are results for the given $\rm LO$, $\rm NLO$, $\rm N^2LO$ and $\rm N^3LO$ series, respectively.}
\label{fig:epsart4}
\end{figure}

\begin{table}[htb]
 	\renewcommand\arraystretch{1.5}
 		\begin{tabular}{cc}
 		\hline
 			&	The predicted $\rm N^4LO$ term of $\eta_A$ under the GB model\\ 
 		\hline
 			Conv.	&$[-0.0049,+0.0075]$\\
 			PMCs	&$[-0.0007,+0.0007]$\\
 			\hline
 		\end{tabular}
 	\caption{The predicted $\rm N^4LO$ term of $\eta_A $ using the GB model under the conventional scale-setting(within the scale range $\mu_{r}\in[Q/2,2Q]$) and the PMCs approaches.}
\label{t5}
\end{table}

To better understand the above results, we also present the probability distribution of $\eta_A$ under the GB model in Fig.\ref{fig:epsart4}. For the case of only LO being known, the distribution manifests as a straight line nearly overlapping the $x$-axis, leading to an infinite standard deviation -- indicating that predictions based on known first-order information alone yield invalid results. When NLO, $\rm N^2LO$, and $\rm N^3LO$ are considered, the functions are all symmetric curves with a central bump, resulting in the mean and median coinciding. As the number of known orders increases, $\eta_A$ becomes increasingly concentrated around the distribution's center, leading to smaller standard deviations and narrower prediction intervals for higher-order terms. Compared to the distributions in Fig.\ref{fig:epsart2}, the GB model's probability distribution function shares many similarities with that of the CH model. The key distinction is that the GB model's central peak is a single point rather than a flat distribution, signifying a value with the highest probability -- where the UHO contribution is zero. Detailed numerical results for the $\rm N^{4}LO$ term of $\eta_A$ are presented in Table~\ref{t5}. 

\begin{table}[htb]
\renewcommand\arraystretch{1.5}
\setlength{\tabcolsep}{12pt}
\begin{tabular}{ccc}
\hline
&$\eta_{A} (\rm Conv.)$&$\eta_{A}(\rm PMCs)$\\ 
\hline
$[0/3]$-type & $0.9265_{-0.0408}^{+0.0458}$&$0.9225_{-0.0162}^{+0.0112}$\\
$[1/2]$-type & $0.9265_{-0.0508}^{+0.0460}$&$0.9225_{-0.0162}^{+0.0112}$\\
$[2/1]$-type & $0.9265_{-0.0508}^{+0.0450}$&$0.9225_{-0.0162}^{+0.0112}$\\
\hline
\end{tabular}
\caption{The values of $\eta_A$ using three types of PAA under the conventional and PMCs approaches.}
\label{t6}
\end{table}

\begin{table}[htb]
\renewcommand\arraystretch{1.5}
\setlength{\tabcolsep}{12pt}
\begin{tabular}{ccc}
\hline
			&$\eta_{A} (\rm Conv.)$&$\eta_{A}(\rm PMCs)$\\ 
\hline
			CH model&$0.9265_{-0.0412}^{+0.0446}$&$0.9225_{-0.0162}^{+0.0113}$\\
			GB model&$0.9265_{-0.0411}^{+0.0451}$&$0.9225_{-0.0162}^{+0.0113}$\\
\hline
\end{tabular}
\caption{The values of $\eta_A$ using the CH and GB Bayesian models under the conventional and PMCs approaches.}
\label{t7}
\end{table}
 
In addition to the uncertainty caused by UHO terms, there is also uncertainty caused by the value of $\alpha_s(M_{_Z})$ in the $\overline{\rm MS}$ scheme. Using $\alpha_s(M_{_Z})=0.1179\pm0.0009$~\cite{ParticleDataGroup:2022pth}, we obtain $\Lambda^V_{{\rm QCD}| n_f=4}=283.0^{+16.1}_{-15.6}\,{\rm MeV}$~\cite{Zhou:2022fyb}. Combining the uncertainties from the renormalization scale, the UHO terms and $\alpha_s(M_{_Z})$, we obtain the final values of $\eta_A$. The numerical results of $\eta_A$ are presented in Table \ref{t6} and Table \ref{t7}. Notably, the numerical results for PMCs appear to be almost the same across both tables. This consistency is due to the fact that the contributions from higher orders of PMCs series are minimal, thereby exerting an almost negligible effect on $\eta_A$.

To facilitate the final numerical analysis, we take the average of the data in Table \ref{t6} and Table \ref{t7} for discussion. Subsequently, we obtain the following results:
\begin{eqnarray}
 	\eta_{A}|_{\rm{Conv.}}^{\rm{PAA}}&=&0.9265^{+0.0456}_{-0.0477},\\
 	\eta_{A}|_{\rm{PMCs}}^{\rm{PAA}}&=&0.9225^{+0.0113}_{-0.0162},\\
 	\eta_{A}|_{\rm{Conv.}}^{\rm{Bayes}}&=&0.9265^{+0.0448}_{-0.0412},\\
 	\eta_{A}|_{\rm{PMCs}}^{\rm{Bayes}}&=&0.9225^{+0.0113}_{-0.0162}.
\end{eqnarray}

\begin{table*}[htb]
\renewcommand\arraystretch{1.5}
\setlength{\tabcolsep}{12pt}
\begin{tabular}{ccccc}
\hline
&$|V_{\rm{{cb}}}||_{\rm{PMCs}}^{\rm{PAA}}$&$|V_{\rm{{cb}}}||_{\rm{Conv.}}^{\rm{PAA}}$&$|V_{\rm{{cb}}}||_{\rm{PMCs}}^{\rm{Bayes}}$&$|V_{\rm{{cb}}}||_{\rm{Conv.}}^{\rm{Bayes}}$\\ 
\hline				
OPAL partial reco \cite{OPAL:2000hcv}  &$ 42.72^{+1.85}_{-1.93}         $&$ 42.83^{+2.76}_{-2.83}            $&$ 42.72^{+1.85}_{-1.93}         $&$    42.83^{+2.73}_{-2.60}         $\\
OPAL excl   \cite{OPAL:2000hcv}      &$ 40.84^{+2.37}_{-2.43}         $&$ 40.94^{+3.08}_{-3.14}            $&$ 40.84^{+2.37}_{-2.43}         $&$    40.94^{+3.05}_{-2.95}         $\\
DELPHI partial reco\cite{DELPHI:2001def} &$ 40.67^{+1.21}_{-1.31}         $&$ 40.77^{+2.29}_{-2.37}            $&$ 40.67^{+1.21}_{-1.31}         $&$    40.77^{+2.26}_{-2.12}         $\\
DELPHI excl    \cite{DELPHI:2004hkn}    &$ 39.94^{+1.19}_{-1.29}         $&$ 40.04^{+2.25}_{-2.33}            $&$ 39.94^{+1.19}_{-1.29}         $&$    40.04^{+2.22}_{-2.08}         $\\
BABAR excl    \cite{BaBar:2007cke}     &$ 39.53^{+1.20}_{-1.30}         $&$ 39.63^{+2.24}_{-2.32}            $&$ 39.53^{+1.20}_{-1.30}         $&$    39.63^{+2.21}_{-2.08}         $\\
BABAR $D^{*0}$   \cite{BaBar:2007nwi}  &$ 40.90^{+1.37}_{-1.47}         $&$ 41.00^{+2.39}_{-2.47}            $&$ 40.90^{+1.37}_{-1.47}         $&$    41.00^{+2.36}_{-2.23}         $\\
BABAR global fit  \cite{ BaBar:2008zui}  &$ 40.44^{+1.98}_{-2.05}         $&$ 40.54^{+2.77}_{-2.84}            $&$ 40.44^{+1.98}_{-2.05}         $&$    40.54^{+2.75}_{-2.64}         $\\
BELLE       \cite{Belle:2018ezy}       &$ 41.92^{+2.19}_{-2.26}         $&$ 42.03^{+2.97}_{-3.04}            $&$ 41.92^{+2.19}_{-2.26}         $&$    42.03^{+2.95}_{-2.84}         $\\
\hline			
\end{tabular}
\caption{The values of $|V_{\rm{{cb}}}|(\times 10^{-3})$ using $\eta_{A}|_{\rm{Conv.}}^{\rm{PAA}}$, $\eta_{A}|_{\rm{PMCs}}^{\rm{PAA}}$, $\eta_{A}|_{\rm{Conv.}}^{\rm{Bayes}}$ and $\eta_{A}|_{\rm{PMCs}}^{\rm{Bayes}}$, which are derived from the data given by various experiment groups~\cite{OPAL:2000hcv,DELPHI:2001def,DELPHI:2004hkn,BaBar:2007cke,BaBar:2007nwi,BaBar:2008zui,Belle:2018ezy} and under CLN parameterization~\cite{Caprini:1997mu}.}
\label{t8}
\end{table*}

Then, using the heavy quark effective theory, $|{\cal F}(1)|=\hat{\xi}(1)\eta_A$, where $\hat{\xi}(1)$ have been calculated in Refs.~\cite{Neubert:1995bc,Gambino:2010bp,Gambino:2012rd}, and $\delta_{1/m_Q^2}$ is about $-(5.5\pm2.5)$$\%$, the value of $|{\cal F}(1)|$ can be obtained. To extract the value of $|V_{\rm{{cb}}}|$, we adopt the Caprini, Lellouch, and Neubert (CLN) parameterization~\cite{Caprini:1997mu} for the form factor ${\cal F}(w)$, and use the experimental data from various experimental groups~\cite{OPAL:2000hcv,DELPHI:2001def,DELPHI:2004hkn,BaBar:2007cke,BaBar:2007nwi,BaBar:2008zui,Belle:2018ezy}. The extracted values of $|V_{\rm{{cb}}}|$ are presented in Table \ref{t8}.

It demonstrates that the results from PMCs approach exhibit smaller theoretical uncertainties and higher precision compared to those from the conventional scale-setting approach. The $|V_{\rm{{cb}}}||_{\rm{PMCs}}^{\rm{PAA}}$ are basically the same as $|V_{\rm{{cb}}}||_{\rm{PMCs}}^{\rm{Bayes}}$. However, by comparing $|V_{\rm{{cb}}}||_{\rm{Conv.}}^{\rm{PAA}}$ and $|V_{\rm{{cb}}}||_{\rm{Conv.}}^{\rm{Bayes}}$, it is evident that the results from the Bayesian models have smaller uncertainties. The weighted average of these theoretical values can be calculated by the formula in Ref.~\cite{ParticleDataGroup:2020ssz}:
\begin{eqnarray}
\overline{x}\pm\delta\overline{x} = \frac{\sum_iw_ix_i}{\sum_iw_i}\pm\left(\sum_iw_i\right)^{-1/2},
\end{eqnarray}
where $\overline{x}$ represents the central value, $\delta\overline{x}$ stands for its uncertainty, and $w_{i}=1/(\delta_{i}x_{i})^{2}$ is the weight factor. Then we obtain the weighted average of $|V_{cb}|$:
\begin{eqnarray}
|V_{\rm{{cb}}}||_{\rm{Conv.}}^{\rm{PAA}}&=&(40.83^{+0.90}_{-0.92})\times10^{-3},\\
|V_{\rm{{cb}}}||_{\rm{PMCs}}^{\rm{PAA}}&=&(40.58^{+0.53}_{-0.57})\times10^{-3},\\
|V_{\rm{{cb}}}||_{\rm{Conv.}}^{\rm{Bayes}}&=&(40.82^{+0.89}_{-0.84})\times10^{-3},\\
|V_{\rm{{cb}}}||_{\rm{PMCs}}^{\rm{Bayes}}&=&(40.58^{+0.53}_{-0.57})\times10^{-3}.
\end{eqnarray}
The above results agree with the PDG average value, i.e., $|V_{\rm{cb}}|_{\rm{PDG}}=(41.1\pm1.2)\times10^{-3}$~\cite{ParticleDataGroup:2024cfk}, within errors. As a comparison, the recent value reported by HFLAV is $|V_{cb}|_{\rm HFLAV} = (38.90 \pm 0.53) \times 10^{-3}$~\cite{HFLAV:2022esi}. This value deviates from the PDG average value by $ 0.89 \sigma$, and deviates from the four predicted values by $0.85 \sigma$, $1.13 \sigma$, $1.00 \sigma$, and $1.13 \sigma$, respectively.

\section{Summary}
\label{sec4}

In this paper, we provide an analysis of the parameter $\eta_A$ for the process $B \to D^{*}\ell \bar{\nu}_{\ell}$ up to the $\rm N^4LO$ level. The $\rm N^4LO$ term of $\eta_A$ was estimated using Bayesian models (the CH and GB models) based on the available pQCD series up to $\rm N^3LO$. For comparison, the $\rm N^4LO$ term was also estimated using the conventional PAA approach. We found that the uncertainties in predictions from the Bayesian models are reduced compared to those from the PAA under the conventional scale-setting approach. However, due to the enhanced convergence of the $\eta_A$ series after applying the PMCs, the predictions of the $\rm N^4LO$ term were largely consistent between the PAA and Bayesian models under the PMCs approach. The results show that the PMCs approach effectively enhances the precision of theoretical predictions by eliminating scale uncertainty relative to the conventional approach. Finally, by comparing the theoretical prediction of the decay width for $B{\rightarrow}D^{*}\ell\bar{\nu}_\ell$ with the latest experimental measurements, we obtained $|V_{\rm cb}|_{\rm PMC} = (40.58^{+0.53}_{-0.57}) \times 10^{-3}$, which is in good agreement with the PDG world average value $|V_{\rm cb}|_{\rm PDG} = (41.1 \pm 1.2) \times 10^{-3}$.

\hspace{2cm}

\noindent {\bf Acknowledgments:} This work was supported by the Natural Science Foundation of China under Grants No.12175025 and No.12347101, by the Research Fund for the Doctoral Program of the Southwest University of Science and Technology under Contract No.24zx7117 and No.23zx7122, and by the Chongqing Natural Science
Foundation under Grant No. CSTB2022NSCQ-MSX0415.

\hspace{2cm}

\end{document}